\documentclass[iop,superscriptaddress,apj]{emulateapj}

\usepackage{hyperref}
\usepackage[usenames]{xcolor}

\newcommand{\nustar}{{\it NuSTAR}}
\newcommand{\chandra}{{\it Chandra}}
\newcommand{\xmm}{{\it XMM-Newton}}
\newcommand{\suzaku}{{\it Suzaku}}
\newcommand\xspec{\textsc{Xspec}}
\newcommand\heasoft{\textsc{HEASoft}}
\newcommand\nustardas{\textsc{NuSTARDAS}}

\newcommand{\keV}{\, \mathrm{keV}}
\newcommand{\Ms}{\,\mathrm{M_\odot}} 					

\newcommand{\erg}{\,\textrm{erg}}
\newcommand{\Mpc}{\,\textrm{Mpc}}
\newcommand{\kpc}{\,\textrm{kpc}}

\newcommand{\cm}{\,\textrm{cm}}
\newcommand{\km}{\,\textrm{km}}

\newcommand\tabref[1]{%
Table~\ref{tab:#1}}

\newcommand\figref[1]{%
Fig.~\ref{fig:#1}}

\newcommand\secref[1]{%
Sec.~\ref{sec:#1}}

\slugcomment{Prepared for ApJ}
\shorttitle{NuSTAR line emission constraints}
\shortauthors{Riemer-S\o{}rensen et al.}

\begin{document}


\title{Dark matter line emission constraints from \nustar{} observations of the Bullet Cluster}


\author{S. Riemer--S\o{}rensen\altaffilmark{1}}

\altaffiltext{1}{Institute of Theoretical Astrophysics, University of Oslo, PO 1029 Blindern, 0315 Oslo, Norway}
\email{signe.riemer-sorensen@astro.uio.no}

\author{D. Wik\altaffilmark{2}} 
\altaffiltext{2}{Astrophysics Science Division, NASA/Goddard Space Flight Center, Greenbelt, MD 20771, USA}

\author{G. Madejski\altaffilmark{3}}
\altaffiltext{3}{Kavli Institute for Particle Astrophysics and Cosmology, SLAC National Accelerator Laboratory, Menlo Park, CA 94025, USA}

\author{S. Molendi\altaffilmark{4}}
\altaffiltext{4}{INAF, IASF Milano, via E. Bassini 15 I-20133 Milano, Italy}

\author{F. Gastaldello\altaffilmark{4}}

\author{F. A. Harrison\altaffilmark{5}}
\altaffiltext{5}{Cahill Center for Astronomy and Astrophysics, California Institute of Technology, Pasadena, CA 91125, USA}

\author{W. W. Craig\altaffilmark{6}\altaffilmark{,7}}
\altaffiltext{6}{Space Sciences Laboratory, University of California, Berkeley, CA 94720, USA}
\altaffiltext{7}{Lawrence Livermore National Laboratory, Livermore, CA 94550, USA}

\author{C. J. Hailey\altaffilmark{8}}
\altaffiltext{8}{Columbia Astrophysics Laboratory, Columbia University, New York NY 10027, USA}

\author{S. E. Boggs\altaffilmark{9}}
\altaffiltext{9}{Space Sciences Laboratory, University of California, Berkeley, CA 94720, USA}

\author{F. E. Christensen\altaffilmark{10}}
\altaffiltext{10}{DTU Space, National Space Institute, Technical University of Denmark, Elektrovej 327, DK-2800 Lyngby, Denmark}

\author{D. Stern\altaffilmark{11}}
\altaffiltext{11}{Jet Propulsion Laboratory, California Institute of Technology, Pasadena, CA 91109, USA}

\author{W. W. Zhang\altaffilmark{2}}

\author{A. Hornstrup\altaffilmark{10}}


\begin{abstract}
Line emission from dark matter is well motivated for some candidates e.g. sterile neutrinos. We present the first search for dark matter line emission in the $3-80\keV$ range in a pointed observation of the Bullet Cluster with \nustar{}. We do not detect any significant line emission and instead we derive upper limits (95\% CL) on the flux, and interpret these constraints in the context of sterile neutrinos and more generic dark matter candidates. \nustar{} does not have the sensitivity to constrain the recently claimed line detection at $3.5\keV$, but improves on the constraints for energies of $10-25\keV$.

\end{abstract}


\section{Introduction} \label{sec:intro}
Dark matter searches are a key pursuit of both astrophysics and particle physics. The scenario where dark matter is in a form of particles which provide gravity, but otherwise interact very weakly with ordinary matter or photons, is the most compelling \citep[e.g.][]{Taoso:2008}. The most promising astrophysical objects for searches for dark matter are clusters of galaxies, along with the Galactic Center and dwarf galaxy satellites to the Milky Way. Here, we consider the top end of the mass scale, galaxy clusters, with total masses - most of it in the form of dark matter - often exceeding $10^{14}\, {\rm M}_\odot \ (\simeq 2 \times 10^{47}$ g). Most cluster mass estimates are inferred from X-ray observations \citep{Vikhlinin:2009,Mantz:2010}. Under the assumption of hydrostatic equilibrium, X-ray data provice a reasonably accurate model of the mass distribution, and imply cluster masses that are roughly consistent with masses measured via gravitational lensing, albeit both types of data are available for only a limited number of objects. Their total masses are roughly five times greater than the baryonic masses inferred from X-ray luminosities \citep{vonderLinden:2014}. 

One possible particle candidate for dark matter is the sterile neutrino \citep[described in the reviews][and references therein]{Boyarsky:2009r,Kusenko:2009,Drewes:2013}. In the framework of this minimally neutrino extension of the standard model ($\nu$MSM), the lightest of the three sterile neutrinos provides the dark matter. The mass is basically unbound from theory, but some astrophysical constraints apply.
The mass is firmly bound from below through the phase space density of nearby dwarf galaxies. The Tremaine-Gunn bound \citep{Tremaine:1979} gives a model independent lower mass of roughly $0.4\keV$ \citep{Boyarsky:2008a}. This limit can be increased if the production method is known; e.g., for resonant production the boundary is approximately $1 \keV$ \citep{Boyarsky:2008a}. An upper limit of a few hundred $\keV$ comes from a combination of production mechanisms and line emission searches. The sterile neutrino can, in principle, decay to two photons of equal energy, and thus the resulting signature would be a narrow emission line, corresponding to an energy of $E_\gamma = m_{\rm s} c^{2}/2$, where $m_{\rm s}$ is the mass of the sterile neutrino. The line width would be determined roughly by the velocity dispersion of dark matter particles, which for clusters is of the order of $v \sim 1000 \km \sec^{-1}$, which is smaller than the instrumental resolution of current observatories. 

If we can limit the X-ray flux of a line at a specific energy, and know the distance, we right away have a limit on the luminosity of the line from the cluster. Since we know the
total mass of the cluster, we know how many sterile neutrinos there must be at a given assumed energy to provide the total mass of the cluster. Since we have a limit on the line luminosity, the ratio of the two is basically the limit on the decay rate. A number of authors have reported searches for such emission lines in the soft X-ray band \citep{Boyarsky:2006a,Boyarsky:2006b,Riemer-Sorensen:2006,Abazajian:2006a,Boyarsky:2007,Riemer-Sorensen:2007,Boyarsky:2008a,Boyarsky:2008b,Riemer-Sorensen:2009,Loewenstein:2009,Loewenstein:2010,Loewenstein:2012}, with a few claims of potential detections that are yet to be confirmed \citep{Loewenstein:2012,Boyarsky:2014,Bulbul:2014}. For an unambiguous detection, the searches must avoid the spectral regions with line emission associated with atomic (or nuclear) transitions from the cluster gas, corresponding to any elements with an expected appreciable cosmic abundance. 

\citet{Bulbul:2014} recently reported a possible signature for such a sterile neutrino at $E_\gamma \simeq 3.5\keV$ in stacked spectra of galaxy clusters observed with the \xmm{} satellite; the result was confirmed at lower significance in a couple of individual clusters \citep{Boyarsky:2014}, but it remains to be independently confirmed in other types of dark matter objects, or using different instruments such as \suzaku{} \citep{Riemer-Sorensen:2014,Tamura:2015,Sekiya:2015,Malyshev:2014,Anderson:2014}. Regardless, since there is no theoretical expectation as to the mass of the dark matter sterile neutrino beyond the broad rage described above, one should search for its signature in all accessible X-ray spectral bands.  

Here, we report on a search extending the energy range to the hard X-ray band, and thus the mass of the putative sterile neutrino to twice the upper end of the \nustar{}'s bandpass, $m_s = 156\keV$. Prior to the launch of \nustar, there were no sensitive spectral measurements beyond $10\keV$, mainly because sensitive measurements require focussing optics \citep{Harrison:2013}. While these energies have been searched for line emission previously using the cosmic background \citep{Boyarsky:2006a,Boyarsky:2008a}, this is the first search in a pointed observation with focussing optics.
  
Recently the \nustar{} team observed and reported the results on one well-studied cluster - the ``Bullet Cluster'' \citep{Wik:2014}. There are many previous observations covering its X-ray and lensing properties \citep{Markevitch:2002,Markevitch:2004,Clowe:2006, Paraficz:2012}. 
This galaxy cluster, at $z = 0.296$, is perhaps best known for the detailed comparison of the distribution of dark matter as inferred from gravitational lensing to the X-ray emitting gas. The lack of spatial alignment between the two distributions reported in \citet{Clowe:2006} on the basis of a weak lensing analysis and in \citet{Bradac:2006} using a joint weak and strong lensing analysis, is often considered to be one of the strongest arguments for the existence of dark matter particles. \citet{Bradac:2006} estimate a total 
mass of the cluster to be $5 \times 10^{14}\,{\rm M}_{\odot}$ within the central $500 \kpc$. 
Due to the offset between the mass and X-ray emitting gas, the Bullet Cluster provides an excellent low-background environment for dark matter searches.

The \chandra{} X-ray data of the Bullet have been searched for isolated X-ray emission lines out to $10\keV$ \citep{Riemer-Sorensen:2007,Boyarsky:2008} but here we extend the range to $\sim 80\keV$ by using the \nustar{} observations from \citet{Wik:2014}. 

Our approach (described in \secref{spectral}) is to fit the data with an adequate model describing the emission by the hot gas in the cluster, and then search for the improvement to the fit by adding an isolated emission line of varying energy but with fixed width (determined by the instrumental resolution). The detection (or limit on the flux of the line) provide the measurement (or limit) of the flux and thus luminosity of decay photons, yielding the volume-integrated decay rate of the putative dark matter particle (presented in \secref{results}).

\section{Observations}
The Bullet Cluster was observed by \nustar{} in two epochs \citep[see][for details]{Wik:2014} for a total combined exposure of 266~ks. To filter the events, standard pipeline processing (\heasoft{} v6.13 and \nustardas{} v1.1.1) was applied along with strict criteria regarding passages through the South Atlantic Anomaly (SAA) and a ``tentacle''-like region of higher activity near part of the SAA; i.e. in the call to  the general processing routine that creates Level 2 data products, {\tt nupipeline}, the following flags were included: {\tt SAAMODE=STRICT} and {\tt TENTACLE=yes}. No strong fluctuations are present in light curves culled from the cleaned events, suggesting a stable background, so no further time periods were excluded.

\begin{figure*}
\centering
\includegraphics[width=0.32\textwidth]{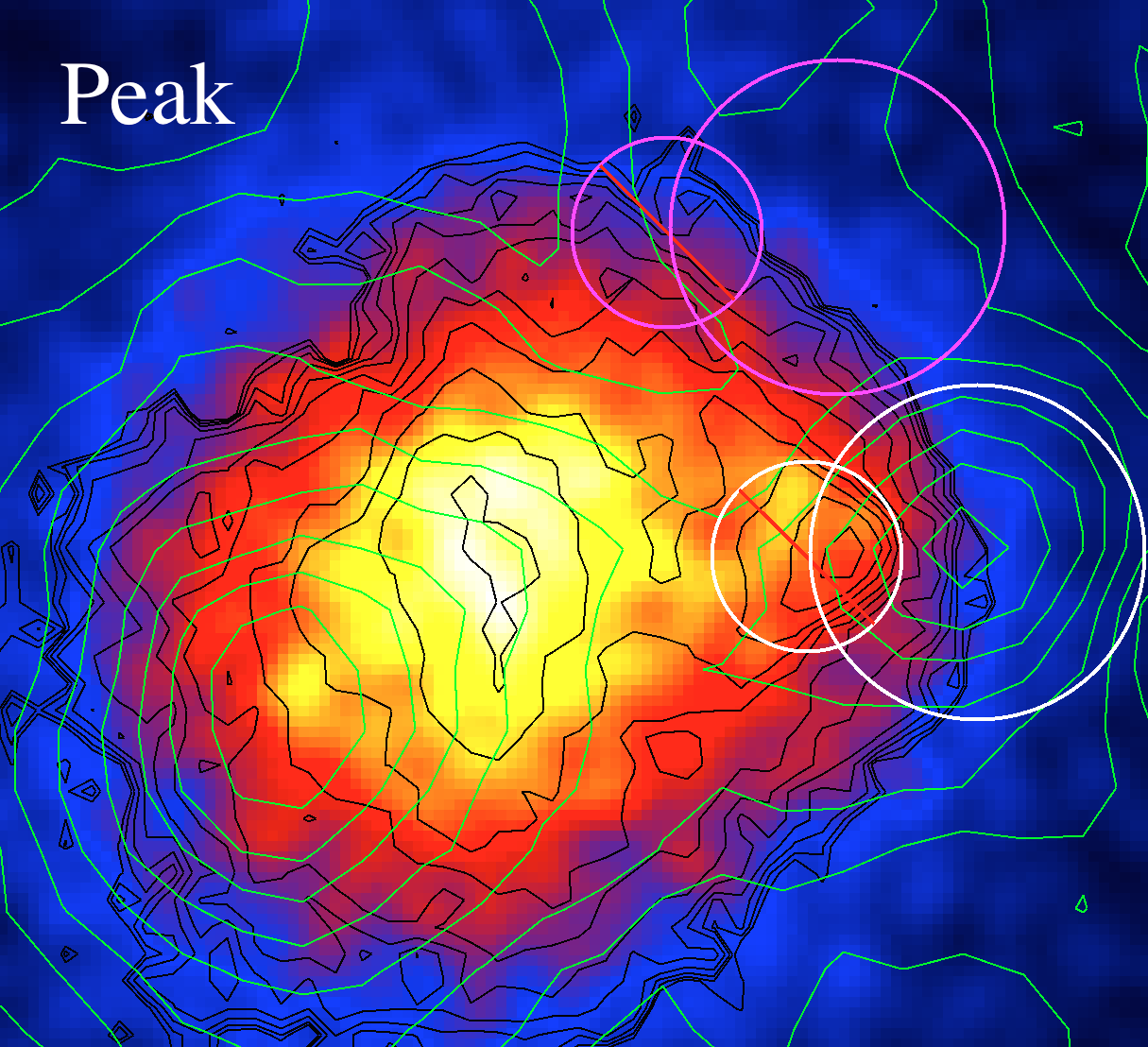}
\includegraphics[width=0.32\textwidth]{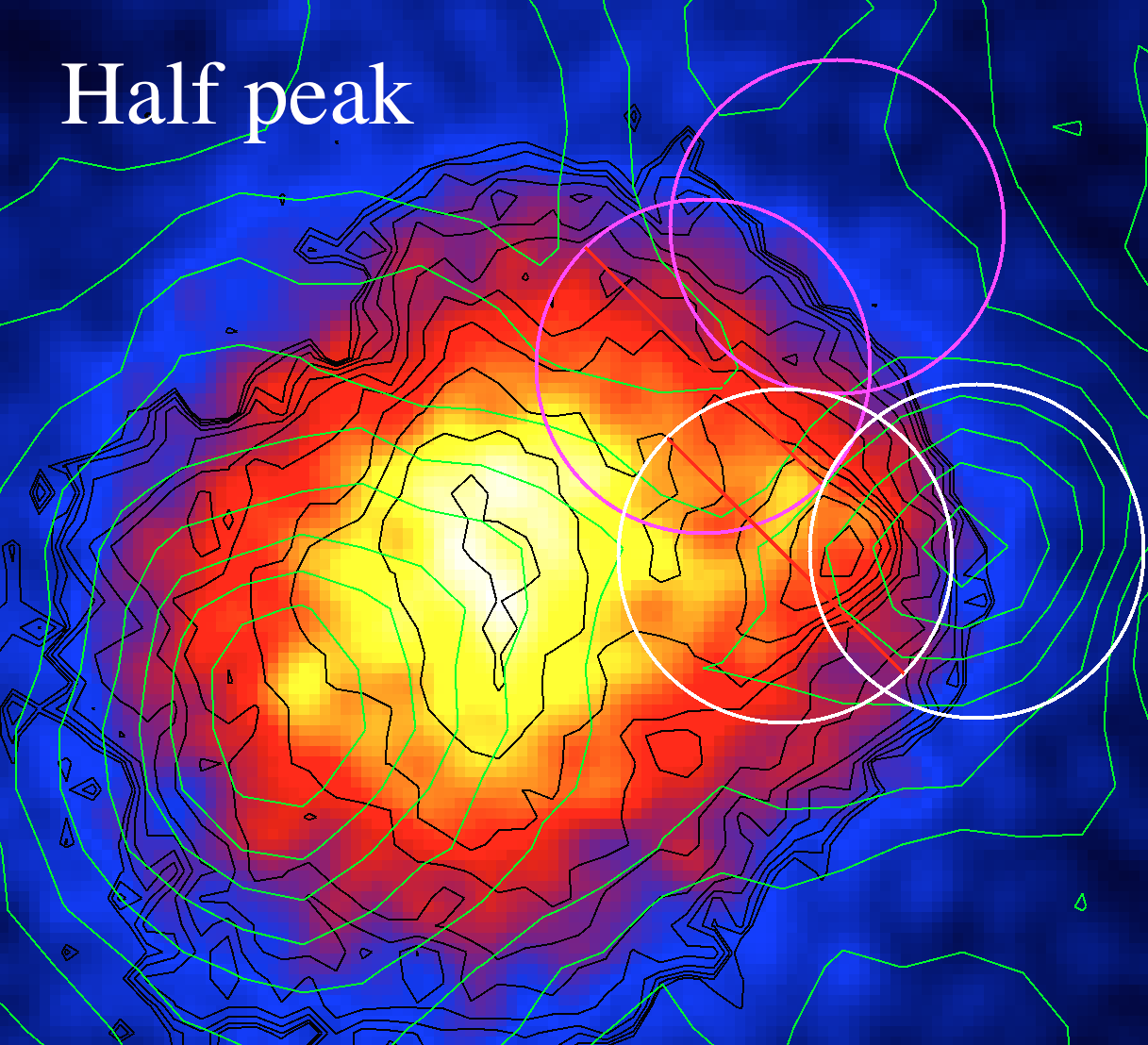}
\includegraphics[width=0.32\textwidth]{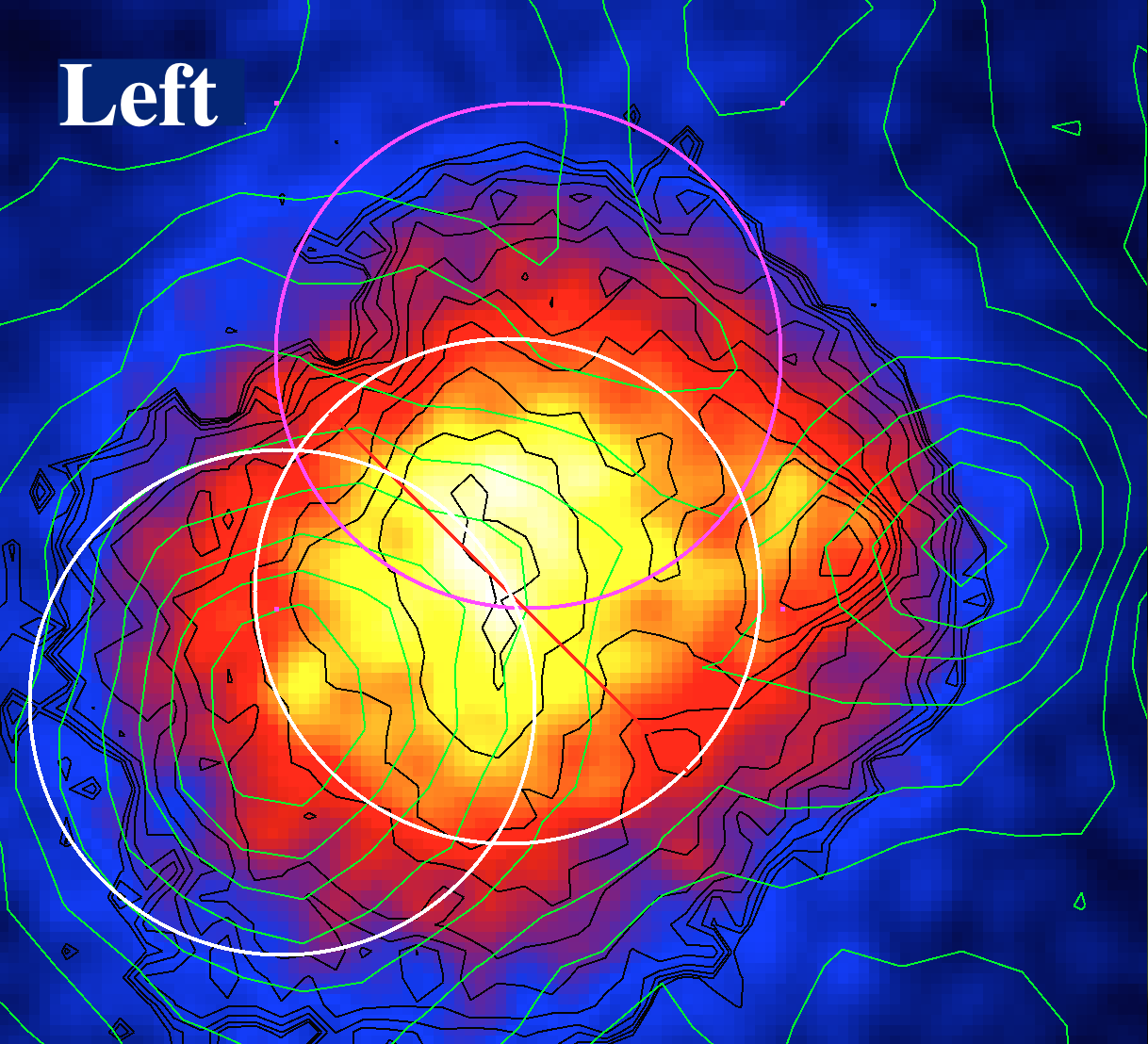}
\caption{Image of the \nustar{} observations of the Bullet Cluster, with the X-ray contours from \chandra{} overlaid in black \citep{Riemer-Sorensen:2007}, and the weak lensing contours in green \citep{Clowe:2006}. The white circles illustrate the source regions with the second circle excluded to avoid gas emission. The magenta circles are the background regions chosen to have similar gas emission as the source regions, but much less mass.}
\label{fig:image}
\end{figure*}

\begin{figure} 
\includegraphics[width=0.49\textwidth]{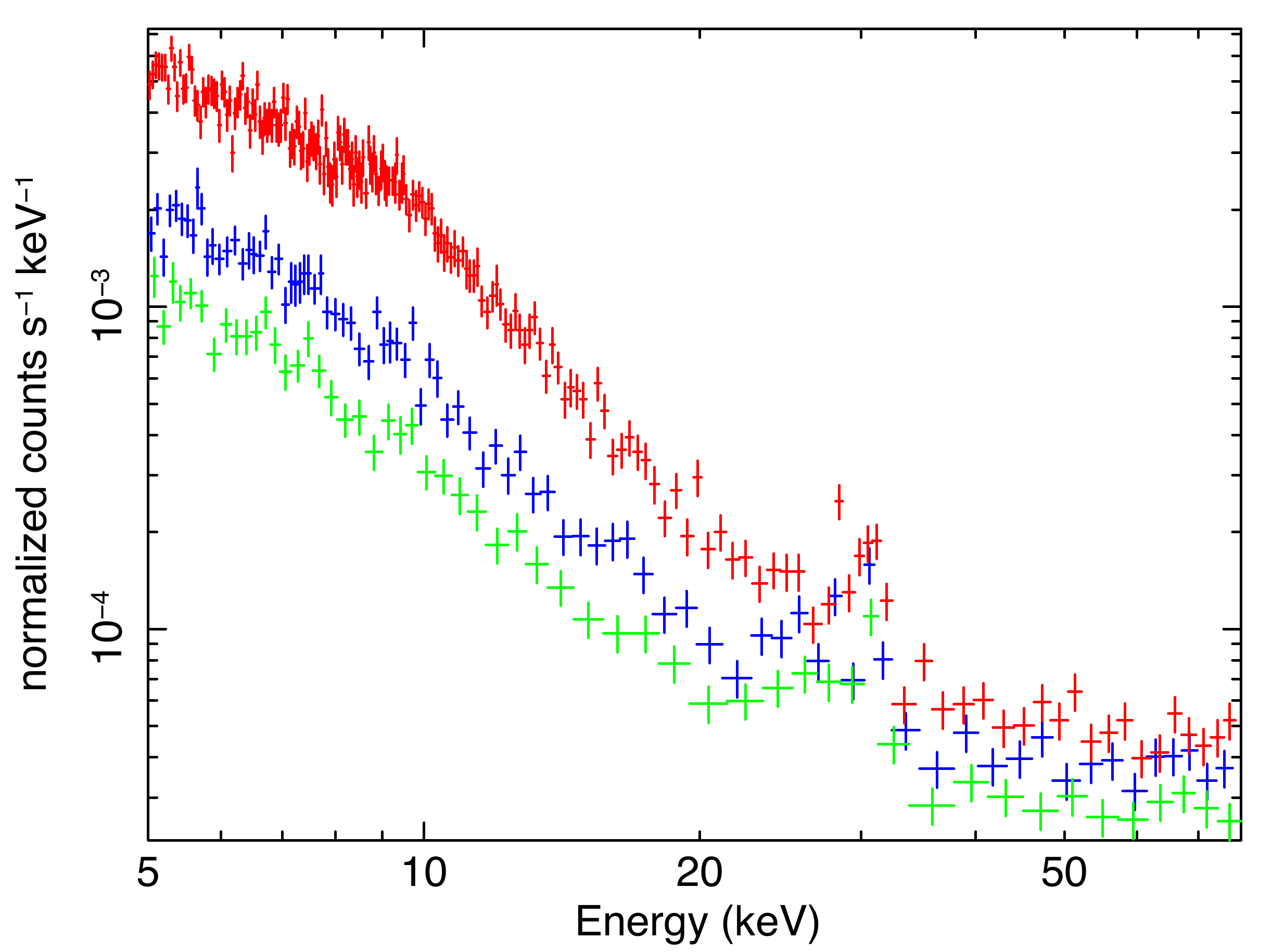}
\caption{The spectra of the three different regions: Peak (blue), Half Peak (green), Left (red). The individual exposures have been stacked and rebinned for visualisation purposes. The bump around $30\keV$ comes from instrumental lines and only affects the background modelling method, not the background subtraction.}
\label{fig:spectra}
\end{figure}

From the cleaned event files, spectra and response files were created using {\tt nuproducts}. The call to {\tt nuproducts} included {\tt extended=yes}, most appropriate for extended sources, which weights the response files based on the distribution of events within the extraction region, assuming that to be equivalent to the true extent of the source.
We extracted spectra for each of the regions shown in \figref{image} (details in \tabref{regions}). The regions were chosen to maximise the amount of dark matter within the field of view while minimising gas emission. The ``Peak'' region contains the leading mass peak but excludes the shock front. It is identical to the analysed region in \citet{Boyarsky:2008} and \citet{Riemer-Sorensen:2007}, while for the ``Half Peak'' region, the extracted area is larger to compensate for the lower spatial resolution of \nustar{} compared to \chandra{}. 
The ``Left'' region contains the trailing mass peak with the exclusion of the main gas emission. For each region we chose an offset background region of identical shape at a location with similar gas emission but less mass as inferred from the lensing map. The resulting spectra for all three regions are shown in \figref{spectra}.

\begin{table*}
\centering
\caption{The first section provides the coordinates of the regions illustrated in \figref{image} and the dark matter mass within each field of view based on weak gravitational lensing. The second section contains the fit statistics for each of the regions for the methods of background subtraction and modelling. The spectra were binned to a minimum of three counts per bin before the analysis.}

\begin{tabular}{l | l | l | l | l}
Region								& Peak					& Half Peak				& Left 				& Peak+Left 				\\ \hline
Included center	(RA, Dec) [degrees]			& 104.56825, -55.941758		& 104.56825, -55.941758		& 104.64978, -55.951826	& --- \\
Included radius [arcmin]					& 0.66					&  0.66					& 1.00				& --- \\
Excluded center (RA, Dec) [degrees] 		& 104.58827, -55.942086		& 104.59071, -55.994209		& 104.62326, -55.944488 	& --- \\
Excluded radius [arcmin]					& 0.375					& 0.66 					& 1.00				& --- \\ \hline
Mass [$\Ms$]							& $5.70\times10^{13}$		& $4.46\times10^{13}$		& $1.02\times10^{14}$	& $1.57\times10^{14}$ \\ \hline

Power law $c$-parameter/dof				& 1460.3/1355				& 1195.6/1105				& 1967.4/1889 			& 3405.2/3221 \\

Background model $c$-parameter/dof		& 1610.0/1646				& 1411.6/1347				& 2380.3/2315			& 3990.3/3933 \\
\end{tabular}
\label{tab:regions}
\end{table*}

\section{Analysis} 
\begin{figure*} 
\centering
\includegraphics[width=0.99\textwidth]{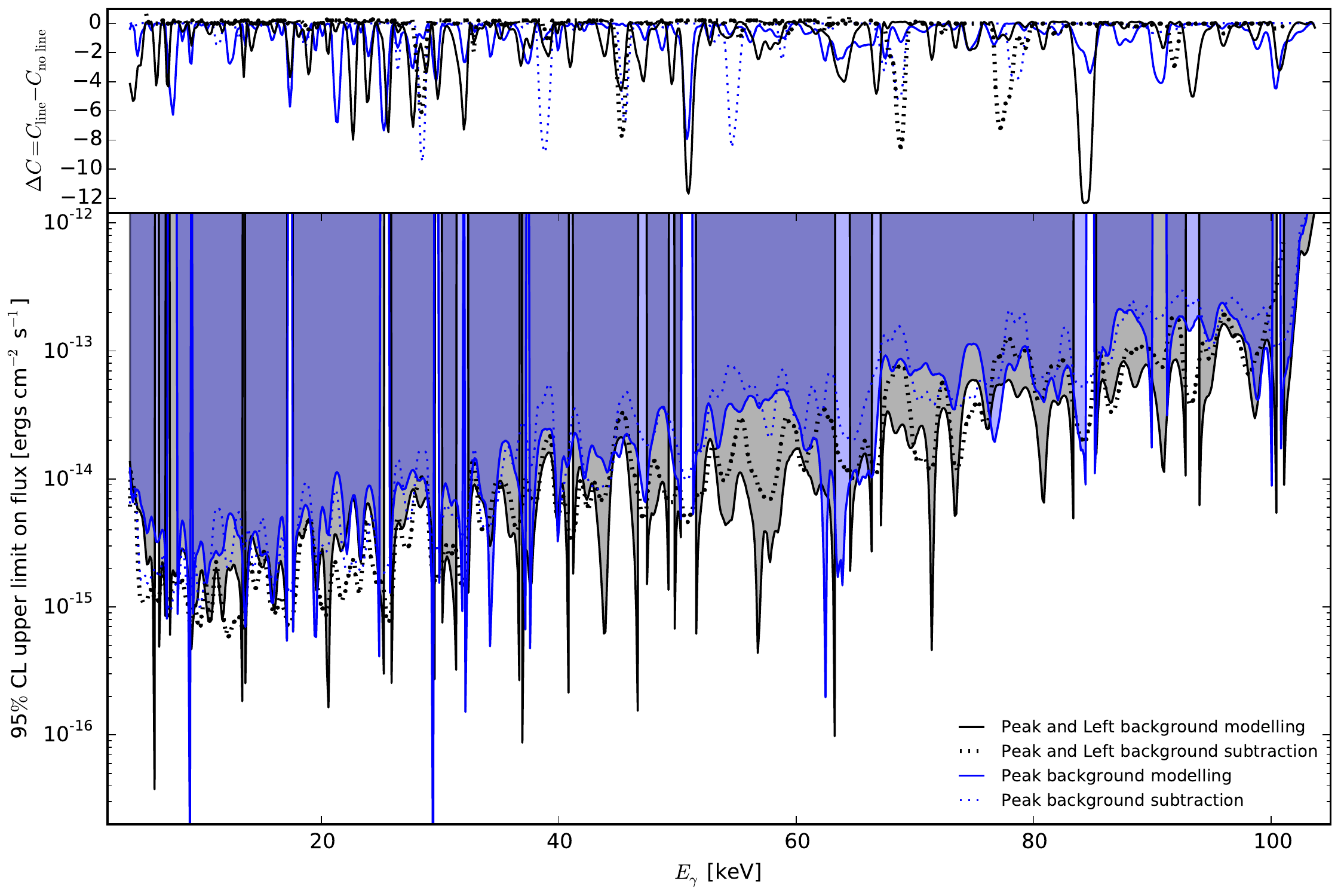}
\caption{{\it Upper panel:} The $C$ statistics improvement (negative) from adding an extra Gaussian at the corresponding rest frame energy ($E_\gamma = (1+z)E_\mathrm{obs}$). {\it Lower panel:} The individual derived flux limits (95\% CL) for the Peak region (blue) as well as the Peak and Left regions combined (thick black). The solid lines show the results of modelling the background of the nearby dark matter offset region, and the dotted lines show the results of subtracting the nearby dark matter offset region. The results from the individual regions are all consistent with each other and the constraints tighten by combining the regions. The background modelling generally provides slightly stronger constraints, but with gaps where the preferred flux is negative.}
\label{fig:flux}
\end{figure*}

\subsection{Spectral modelling} \label{sec:spectral}
We fit the spectra of both observations and both detectors simultaneously using the \xspec{} spectral fitting package \citep{Arnaud:1996} and explored two different approaches to background modelling and subtraction. The spectra were binned to contain at least three counts per bin, and we assume the background to be Poisson distributed and consequently use Cash-statistics ($C$) to optimise the parameter values \citep{Cash:1979}.

In the first method of background treatment we simply subtract the spectrum of the dark matter offset region from the spectrum of the source region. Since the spectra are almost identical we can fit any residuals with a single power law (the statistics for each region are given in \tabref{regions}). Subsequently, we added a Gaussian to represent a single emission line at a fixed energy and flux to the best fit model above, and searched for the improvement of $C$ as a function of line energy and intensity, allowing for simultaneous variation of the power law parameters and the Gaussian normalisation. We consider line intensities from 0 to $10^{-5}\, \mathrm{photons}\cm^{-2}\sec^{-1}$ and line energies of 3 to 80 keV in steps of $\Delta E = 0.1\keV$. We assume that the line is narrow compared to the detector resolution, fixing the intrinsic line width at $0.001 \keV$, and noting that as long as the assumed width is less than $\sim 0.03 \keV$, our results do not change. 

The reduction of the $C$ parameter by the extra line is shown as a function of line energy in the upper panel of \figref{flux}. At most energies, the additional Gaussian does not lead to any significant improvement  ($|\Delta C|<9$ for 1105-3821 degrees of freedom), and instead we constrain the flux by increasing the Gaussian normalization and refitting all other parameters until $\Delta C = C - C_\mathrm{base} = 2.71$, corresponding to the one-sided $95\%$ confidence level marginalised over the power law normalizations. These flux levels are shown in \figref{flux} for the Peak and Left regions as well as for the combined analysis.


In the second approach we model the background instead of subtracting it. In \citet{Wik:2014} the background emission in the Bullet Cluster was thoroughly investigated and we use their results as a baseline model for the background, and check if there is room for any line emission above this model. The model consists of four components:
i) the aperture background (smooth gradient across the detector with a normalisation uncertainty of 10\%); 
ii) a focused cosmic ray background from unresolved sources (smooth with a normalisation uncertainty of 10\%); and 
iii) instrumental continuum (we use a 10\% normalisation uncertainty even though the systematic uncertainty is probably much smaller); 
iv) the thermal solar continuum and instrumental lines from reflections (smooth component with a normalisation uncertainty of 10\% plus known detector emission lines). 
Additionally we fit a line free plasma model \cite[apec,][]{Smith:2001} to account for any gas emission from the cluster. The redshift and abundances of the plasma model are kept fixed at $z=0.296$ and $A=0.2$ respectively (consistent with the value found in \citep{Wik:2014}), while the temperature is required to be the same for both detectors and both observations, but with individual normalisations.

The $3-150\keV$ energy interval is well fitted by the background model with the statistics given in \tabref{regions}. As before, for each energy between $3$ and $80\keV$ in steps of $\Delta E_\gamma = 0.1\keV$ we add a Gaussian and determine the best fit normalisation considering line intensities from $-10^{-5}$ to $10^{-5}\, \mathrm{photons}\cm^{-2}\sec^{-1}$. The line intensities are allowed to be negative to account for overestimation of the background. 
The normalisation is then increased until $\Delta C^2=2.71$. We derive upper flux limits from the Gaussian alone as well as from the base model plus the Gaussian integrated over the full width at half maximum (FWHM) of the \nustar{} spectral resolution crudely approximated by \citep{Harrison:2013}
\begin{equation}
\Delta E_{\rm FWHM} = 0.01E_\gamma + 0.3 \keV \, .
\end{equation}
The addition of the Gaussian only improves the model by $|\Delta C| > 9$ around $50$ and $84 \keV$ where $\Delta C \approx -12$, reflected in weaker constraints at those energies. 
While $|\Delta| C > 9$ would appear to imply a significant detection, we need to take the look-elsewhere effect into account \citep{Gross:2010}. The energy of the line is unknown and by scanning over energy we perform a number of independent searches which increases the chance of seeing statistical outliers. Consequently the probability of detection is degraded by the number of independent attempts (given the spectra resolution of \nustar{} we search of the order of 150 independent energies).



Line-like features may arise from fluorescent and activation induced instrumental lines if imperfectly modelled, or due to statistical/systematic fluctuations between target and background regions \citep[see appendix of][]{Wik:2014}. These lines are strongest between $20-30\keV$, and may explain the region of larger $C$-values at those energies.

The flux limits from the two approaches for the background subtraction are compared in \figref{flux}. The solid lines show the results of modelling the background of the nearby dark matter offset region, and the dotted lines show the results of subtracting the nearby dark matter offset region. The results from the individual regions are consistent with each other, and the constraints tighten by combining the regions. The background modelling generally provides slightly stronger constraints, but with gaps where the preferred flux is negative. The results are very similar, and we consider the background treatment to be robust. We also derive flux limits from fitting the two regions simultaneously, taking their mass difference and consequently expected signal strength into account.

While the background modelling provide slightly stronger constraints, there is a risk of spurious line detections from small residuals due to a gain drift between the actual observation and the time of collection of data from which the background model was constructed. This is unlikely to be the case for direct background subtraction.

\subsection{Mass within field of view}
The weak lensing shear maps\footnote{\url{http://flamingos.astro.ufl.edu/1e0657/index.html}} from \citet{Clowe:2006} can be integrated to provide the masses within the three regions for a fiducial cosmology of $\Omega_\mathrm{m}=0.3$, $\Omega_\Lambda = 0.7$, $H_0 = 70 \km\sec^{-1}\Mpc^{-1}$. The map contours are shown in \figref{image} and the obtained region masses are given in \tabref{regions}.

\citet{Paraficz:2012} presented a mass map of the Bullet Cluster based on strong lensing rather than weak lensing. This map provides region masses that are almost twice as big as for the weak lensing map. 
This indicates an uncertainty on the mass estimates of the order of 50\%. In the remaining analysis and plots we conservatively use the values from the map by \citet{Clowe:2006} and assume that the entire mass is made up of dark matter since the observed regions have been chosen to minimise the gas presence, so for the Peak region $M_\mathrm{gas} \approx 0.1 M_\mathrm{tot}$.

\section{Results} \label{sec:results}

\begin{figure*} \label{fig:msin}
\centering
\includegraphics[width=0.99\textwidth]{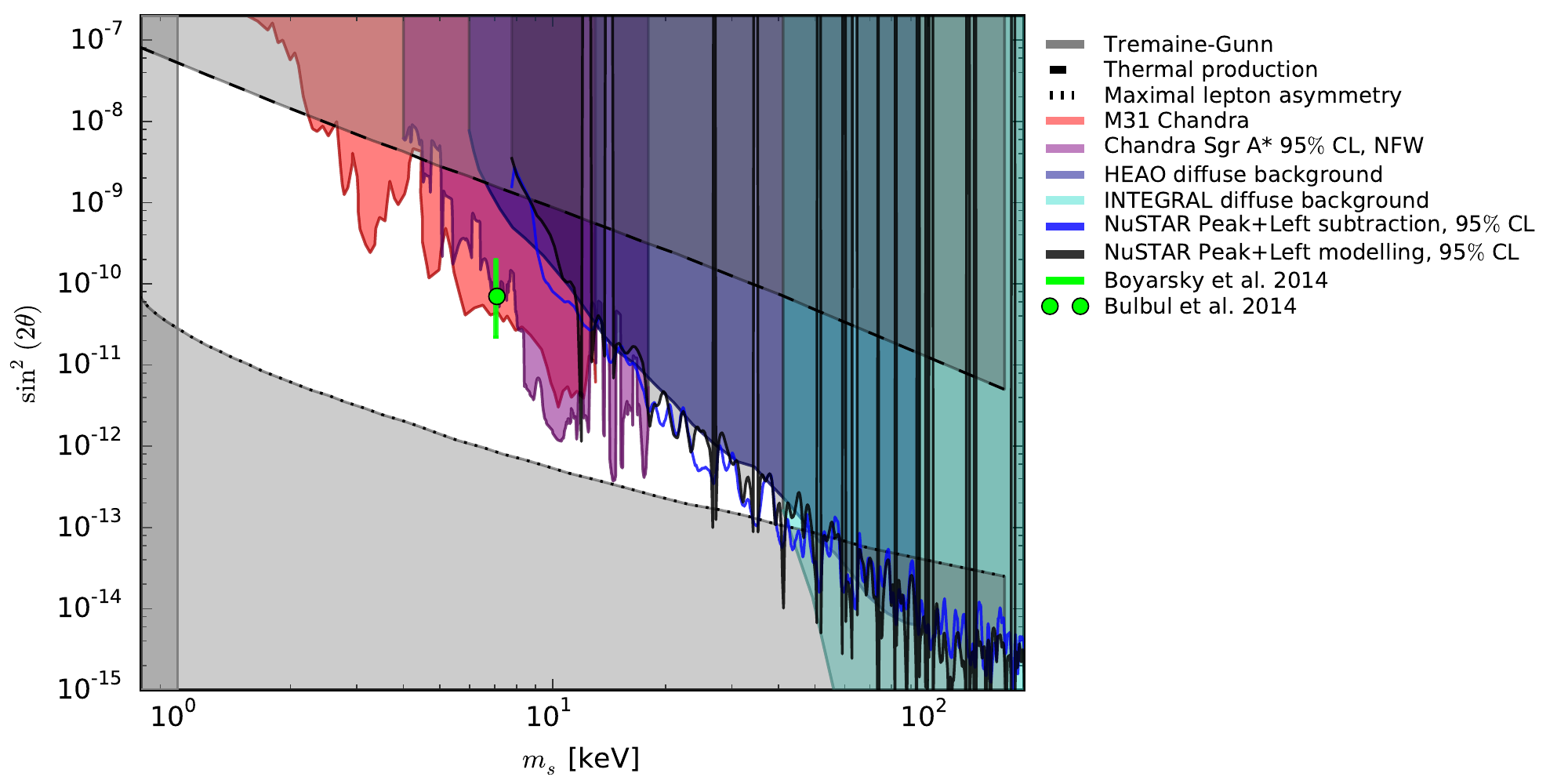}
\caption{The derived constraints on sterile neutrino mixing angle as a function of mass compared to a selection of previous constraints. The upper grey region is excluded by over-production of dark matter, and the lower by Big Bang nucleosynthesis \citep{Laine:2008a,Canetti:2013}. Below $m_s = 1\keV$ the Tremaine-Gunn limit applies \citep{Tremaine:1979, Boyarsky:2008a}. The green points indicate the potential signal from \citep{Bulbul:2014} and \citep{Boyarsky:2014} and the coloured shaded regions are observation bounds: 
M31 Chandra \citep{Horiuchi:2014},
Chandra Sgr A* \citep{Riemer-Sorensen:2014},
HEAO diffuse background \citep{Boyarsky:2006a},
INTEGRAL diffuse background \citep{Boyarsky:2008a}.
The \nustar{} Bullet Cluster constraints are similar to previous constraints from the diffuse cosmic background, but provide an important cross-check.}
\label{fig:msin}
\end{figure*}

\begin{figure*} \label{fig:gamma}
\centering
\includegraphics[width=0.99\textwidth]{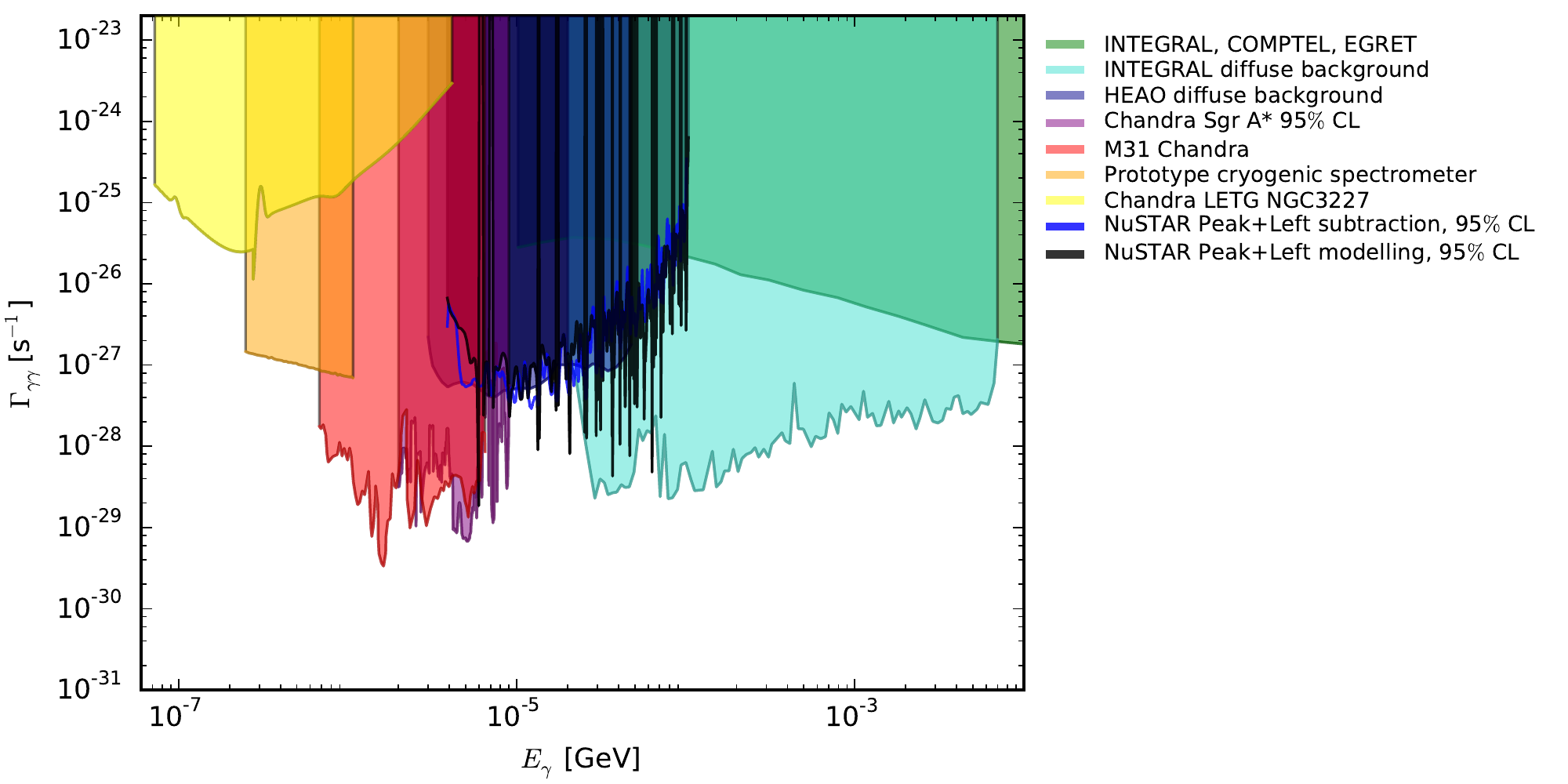}
\caption{The derived constraints on two-photon decay rates as a function of photon energy compared to previous constraints:
INTEGRAL, COMPTEL and EGRET \citep{Yuksel:2008b},
INTEGRAL diffuse background \citep{Boyarsky:2008a},
HEAO diffuse background \citep{Boyarsky:2006a},
Chandra Sgr A* \citep{Riemer-Sorensen:2014},
M31 Chandra \citep{Horiuchi:2014},
prototype cryogenic spectrometer \citep{Boyarsky:2007b},
Chandra LETG NGC3227 \citep{Bazzochi:2008}. 
These constraints applies to all Majorana type dark matter candidates with mono-energetic photon emission in the relevant energy range. The \nustar{} Bullet Cluster constraints are similar to previous constraints from the diffuse cosmic background, but provide an important cross-check based on different assumptions.}
\label{fig:Egamma}
\end{figure*}

\subsection{Sterile neutrinos}
Assuming all of the dark matter to be sterile neutrinos, we can interpret the flux constraints from \figref{flux} in terms of the sterile neutrino parameters of mass, $m_s$ and mixing angle, $\sin^2(2\theta)$, where the latter describes the mixing probability with the lightest of the active neutrinos. The constraints are converted as \citep{Riemer-Sorensen:2006, Boyarsky:2007}:
\begin{eqnarray}\label{eq-massmixing}
&& \sin^2(2\theta) \leq \nonumber \\
&& 10^{18} \left(\frac{F_\mathrm{obs}}{\erg\cm^{-2}\sec^{-1}}\right) \left( \frac{m_s}{\keV}\right) \left[ \frac{(M_\mathrm{fov}/\Ms)}{(D_L/\Mpc)^2} \right]
\end{eqnarray}
where $F_\mathrm{obs}$ is the observed flux limit, $M_\mathrm{fov}$ is the total dark matter mass within the field of view, and $D_{L}$ is the luminosity distance.

As illustrated in \figref{msin}, very large mixing angles will lead to over-production of dark matter and are consequently ruled out. Similarly, the resonant production mechanisms require a primordial lepton asymmetry \citep{Boyarsky:2008a}, that may affect Big Bang nucleosynthesis and cosmic element abundances. This gives a lower limit on the mixing angle. As mentioned in \secref{intro}, the mass range is limited from below by the Tremaine-Gunn bound \citep{Tremaine:1979,Boyarsky:2008a} and from above by line emission searches. The \nustar{} Bullet Cluster constraints from the Peak region alone are weaker than previously existing constraints from the diffuse cosmic background, but by combining the Peak and Left regions we get similar constraints. This provides an important independent cross-check of the many assumptions about, e.g., source distribution that goes into the diffuse background constraints.

\subsection{Generic dark matter constraints}
In \figref{Egamma} we present the constraints on generic dark matter decays leading to photon emission for two photons per decay. For one-photon interactions, the constraints are weaker by a factor of two. For particles of Majorana type where the particles are their own anti-particles, the interaction probability doubles and the constraints are strengthened by a factor of two.

Again, the \nustar{} constraints do not significantly improve on existing constraints, but being a pointed observation with robust background treatment they provide an important cross-check on previous analysis of diffuse emission observed with HEAO, INTEGRAL, COMPTEL and EGRET \citep{Yuksel:2008b, Boyarsky:2008a,Boyarsky:2006a}.

\section{Discussion}

\subsection{Possible detections}
Unfortunately, the claims of possible line detections \citep{Loewenstein:2010, Bulbul:2014, Boyarsky:2014} all lie below the lower sensitivity cutoff for \nustar{} of $3.89\keV$ introduced by the redshift of the Bullet Cluster. 

\subsection{Possible improvements}
The Milky Way halo has provided strong constraints on line emission in the X-ray range \citep{Riemer-Sorensen:2006,Riemer-Sorensen:2014}. The advantage of local constraints is that the dark matter source is nearby and there exists a wealth of observations, but the disadvantage is the number of sources of non-thermal ``background'' radiation and the uncertainty of the inner mass profile. The background radiation issue can be reduced significantly by point source removal if one has sufficient spatial resolution. This is now becoming possible with \nustar{}, and will be investigated further in future work. The profile uncertainty problem can be mitigated somewhat by excluding the centre of the halo from the analysis. 

\section{Summary}
We have searched \nustar{} observations of the Bullet Cluster for exotic line emission over the $3-80\keV$ interval. No significant line flux was found and we have derived upper limits on the possible line emission flux. While the constraints are similar to previous constraints from the cosmic background emission, this is the first time a search has been performed in this energy interval using a pointed observation. The constraints can be improved by longer observations or using different targets, e.g., the Milky Way halo \citep[as in][]{Riemer-Sorensen:2014}.

\acknowledgments
This research made use of data from the \nustar{} mission, a project led by the California Institute of Technology, managed by the Jet Propulsion Laboratory, and funded by NASA, and it also made use of the \nustar{} Data Analysis Software (\nustardas) jointly developed by the ASI Science Data Center (ASDC, Italy) and the California Institute of Technology (USA).
{\it Facilities:} \facility{\nustar{}}.

\bibliographystyle{apj}

\bibliography{sterile}


\end{document}